\documentstyle[12pt,epsf,epsfig]{article}
\newcount\timecount
\newcount\hours \newcount\minutes  \newcount\temp \newcount\pmhours

\hours = \time
\divide\hours by 60
\temp = \hours
\multiply\temp by 60
\minutes = \time
\advance\minutes by -\temp
\def\hour{\the\hours}
\def\minute{\ifnum\minutes<10 0\the\minutes
            \else\the\minutes\fi}
\def\clock{
\ifnum\hours=0 12:\minute\ AM
\else\ifnum\hours<12 \hour:\minute\ AM
      \else\ifnum\hours=12 12:\minute\ PM
            \else\ifnum\hours>12
                 \pmhours=\hours
                 \advance\pmhours by -12
                 \the\pmhours:\minute\ PM
                 \fi
            \fi
      \fi
\fi
}

\def\monthname{\relax\ifcase\month 0/\or January\or February\or
   March\or April\or May\or June\or July\or August\or September\or
   October\or November\or December\else\number\month/\fi}

\def\bold#1{\setbox0=\hbox{$#1$}%
     \kern-.025em\copy0\kern-\wd0
     \kern.05em\copy0\kern-\wd0
     \kern-.025em\raise.0433em\box0 }

\def\ga{\mathrel{\raise.3ex\hbox{$>$\kern-.75em\lower1ex\hbox{$\sim$}}}}
\def\la{\mathrel{\raise.3ex\hbox{$<$\kern-.75em\lower1ex\hbox{$\sim$}}}}
\def\gev{{\rm \, Ge\kern-0.125em V}}
\def\tev{{\rm \, Te\kern-0.125em V}}
\def\beq{\begin{equation}}
\def\eeq{\end{equation}}

\def\mchi{m_{\chi}}
\def\ohsq{\Omega_{\chi} h^2}

\def\m12{m_{1\!/2}}

\begin{document}
\begin{titlepage}
\pagestyle{empty}
\baselineskip=21pt
\rightline{hep-ph/9705444}
\rightline{CERN--TH/97-105}
\rightline{UMN--TH--1537/97}
\rightline{TPI--MINN--97/16}
\vskip 0.2in
\begin{center}
{\large{\bf
Constraints on Neutralino Dark Matter \\
from \\
LEP~2 and Cosmology}} \\
\end{center}
\begin{center}
\vskip 0.2in
{John Ellis$^1$, Toby Falk,$^2$ Keith A.~Olive,$^2$
and Michael Schmitt$^3$
}\\
\vskip 0.1in
{\it
$^1${TH Division, CERN, Geneva, Switzerland}\\
$^2${School of Physics and Astronomy,
University of Minnesota, Minneapolis, MN 55455, USA}\\
$^3${PPE Division, CERN, Geneva, Switzerland}\\}
\vskip 0.2in
{\bf Abstract}
\end{center}
\baselineskip=18pt \noindent
%%%%%%%%%%%%%%%%%%%%%%%%%%%%%%%%%%%%%%%%%%%%%%%%%%%%%%%%%%%%%%%%%%%%%
A significant lower limit on the mass of the lightest neutralino
$\chi$ can be obtained by combining the results from sparticle
searches at LEP at centre-of-mass energies up to 172 GeV with
cosmological considerations, if it is assumed that the $\chi$ is
stable.  Exclusion domains from slepton searches close $m_{\chi} \sim
0$ loopholes that were left open by previous lower-energy LEP searches
for charginos and neutralinos, leading to the lower limit $m_{\chi}
\ga 17$ GeV. The constraints on supersymmetric parameter space are 
strengthened significantly if LEP
constraints on supersymmetric Higgs bosons are taken into account,
and further if the relic neutralino density is
required to fall within the range favoured by astrophysics and cosmology.
These bounds are considerably
strengthened if universality at the GUT scale
is assumed for soft supersymmetry-breaking scalar masses,
including those of the Higgs bosons.
In this case, the Higgs searches play a dramatic r\^ole, and we
find that $m_{\chi} \ga 40$ GeV.
Furthermore, we find that if tan$\beta \la 1.7$ for $\mu<0$, or
tan$\beta \la 1.4$ for $\mu >0$, the cosmological
relic density is too large for all values of $m_{\chi}$.
%%%%%%%%%%%%%%%%%%%%%%%%%%%%%%%%%%%%%%%%%%%%%%%%%%%%%%%%%%%%%%%%%%%%%
\vfill
\leftline{CERN--TH/97-105}
\leftline{May 1997}
\end{titlepage}
\baselineskip=18pt
%%%%%%%%%%%%%%%%%%%%%%%%%%%%%%%%%%%%%%%%%%%%%%%%%%%%%%%%%%%%%%%%%%%%%
\section{Introduction}

Among the most significant and model-independent accelerator
constraints on supersymmetric dark matter candidates are
those provided by LEP, thanks to its very clean experimental
conditions. Many analyses have been conducted within the context of the
minimal supersymmetric
extension of the Standard Model (MSSM) with soft
supersymmetry-breaking parameters - scalar masses $m_i$,
gaugino masses $M_{\alpha}$ and trilinear couplings $A_{ijk}$ - 
originating at some high supergravity scale and evolved down to
lower energies using the renormalization group~\cite{mssm}. Further,
it is often assumed that $R$ parity is
conserved, so that the lightest supersymmetric particle is
stable, and often taken to be
the lightest neutralino $\chi$~\cite{ehnos}. Within this
framework, the negative results
of LEP 1 searches for $Z^0 \rightarrow \chi^+ \chi^-$ (where $\chi^{\pm}$
denotes the lightest chargino) and $Z^0 \rightarrow \chi \chi'$
(where $\chi'$ denotes a generic heavier neutralino) already
established important limits on supersymmetric model parameters,
but left open the possibility that the lightest neutralino might
be massless~\cite{LEP1}.
Subsequently, the advent of data from
higher-energy
LEP runs at energies between 130 and 140 GeV (called here the LEP~1.5
run)~\cite{LEP15}
complemented LEP~1 data in an elegant manner that almost excluded
the possibility of a massless neutralino, at least if the input
gaugino masses $M_{\alpha}$ were assumed to be universal~\cite{achi}.

With these same assumptions, we showed in a previous paper~\cite{efos}
that the remaining loopholes in the experimental analysis could be
blocked, and interesting lower limits on $m_{\chi}$ obtained, by
combining LEP data with those from other $e^+e^-$ experiments~\cite{AMY}.
These
bounds could be strengthened by assuming universality between the
input slepton and squark masses and imposing the cosmological
requirement that the relic neutralino density fall within the
interesting range. We also found that the lower bound on $m_{\chi}$
could be further improved if one extended the assumption of universal
soft supersymmetry-breaking scalar masses $m_i$ to the Higgs sector,
in the context of an implementation of dynamical electroweak symmetry
breaking (EWSB)~\cite{cmssm}.

As the centre-of-mass energy of LEP is increased in steps, this type
of lower bound on the neutralino mass can be strengthened
progressively, and we have already commented\cite{efosp} on the
potential improvement that could be obtained by taking into account
the data produced by LEP close to the $W^+W^-$ threshold at $E_{CM} =
161$ GeV (called here the LEP~2W run)~\cite{LEP2W}.  Recently, results
have been
announced from the subsequent higher-energy run at $E_{CM} = 170/172$
GeV (LEP~2)~\cite{LEP2}, and the main purpose of this paper is to consider
their
implications for the MSSM parameter space and supersymmetric dark
matter [for a recent analysis of the chargino and cosmology
constraints, see~\cite{CK}]. The LEP~2 data of particular interest to us
are the searches
for charginos and neutralinos, new lower limits on the masses of
sleptons $\tilde \ell$, improved limits on the production of stop
squarks $\tilde t$~\cite{stopLEP}, and upper limits on the
production rates for the
neutral supersymmetric Higgs bosons $h,A$. The latter constraints are
especially important when combined with cosmology and/or Higgs mass
universality.

We now find, in contrast to the previous LEP~1.5 analysis, that the
experimental searches for $\chi^+ \chi^-, \chi \chi'$ and $\bar{\tilde
\ell} {\tilde \ell}$
production at LEP~2 leave no loopholes for massless or light
neutralinos, even in the absence of any other phenomenological
inputs apart from gaugino mass universality. This is because 
slepton searches restrict the possibility for charginos to decay
undetected into soft leptons. 
As shown in Fig.~1, we find a lower bound $m_{\chi} \ga 17$~GeV,
if the input slepton masses are assumed to be universal.
Significant extra domains of supersymmetric parameter space are
excluded if one takes
into account the
negative results of LEP searches for supersymmetric Higgs bosons, 
and assumes that all the input slepton and squark masses are
universal, and also
if one assumes that the relic neutralino density
$\Omega_{\chi}$ is large enough to be of astrophysical interest, but
does not overclose the Universe: $0.1 \le \Omega_{\chi} h^2 \le 0.3$.
The lower bound on $m_{\chi}$ may be strengthened significantly if
the universality assumption is extended to the
masses of the Higgs bosons that are
put into renormalization-group
calculations that implement dynamical EWSB, in which case
LEP Higgs searches play a more important r\^ole~\footnote{This point was
recognized implicitly in~\cite{BB}, where it was noted in section IV
that LEP should discover a Higgs boson if tan$\beta$ is small and $\mu <
0$.}.
Within this wholly universal framework, we find the
lower limit $m_{\chi} \ga 40$ GeV, attained when 
$\mu < 0$ and tan$\beta = 2.8$.
This constraint is considerably stronger than that inferred indirectly
from unsuccessful squark and gluino searches by the CDF and D0
collaborations~\cite{gluino}. For small tan$\beta$, the Higgs constraints
improve the lower limit on $m_{\chi}$ so dramatically
that it becomes incompatible
with the cosmological upper limit on the relic density $\Omega_{\chi} h^2 
\le 0.3$. Thus cosmology and LEP~2 data together require the lower limit
tan$\beta \ga 1.7$ for $\mu < 0$ and tan$\beta \ga 1.4$ for $\mu > 0$. 
In
passing, we point out that the LEP~2 results
exclude a large fraction of the domain of parameter space where the
neutralino is a higgsino, and that future higher-energy LEP runs will
be able to determine the fate of this option.

\section{Review of Accelerator Constraints}

As a prelude to our analysis, we first summarize the most relevant
LEP~2 constraints that we use~\cite{LEP2}. The unsuccessful searches for
$e^+e^- \rightarrow \chi^+ \chi^-$ production impose an upper limit on its
cross section $\sigma_{+-}$, which we conservatively estimate as 
$\sigma_{+-} < 0.35$~pb in the regions of parameter space relevent 
to our limit on $m_\chi$,
except when the sneutrino is lighter than the
chargino, in which case
we assume zero efficiency,
which certainly is true when $|m_{\chi^{\pm}} - m_{\tilde \nu}| <
3$~GeV~\footnote{We note that, although the experimental
efficiency decreases
in the limit $m_{1/2} \gg \mu$  where the mass difference 
$m_{\chi^{\pm}} - m_\chi$ becomes small, this case is not relevant 
for this study.}.
There is also an
upper limit on the cross section for associated $\chi \chi'$
production, but this is does not exclude a significant extra domain of
the MSSM parameter space.  In addition to the constraints imposed by
the chargino searches, we find that a useful r\^ole is also played by
the upper limit on selectron~\footnote{The limits on $\tilde \mu$ and
$\tilde \tau$ production do not significantly strengthen the bounds obtained 
from $\tilde e$ alone.}
pair production,
conservatively estimated using a rough approximation for the experimental
efficiencies and the number of reported candidates:
no limit is assumed when $m_{\tilde e}-\mchi < 10\gev$.
Among the LEP~2 constraints of most interest to
us are those on supersymmetric Higgs production, for which we consider
both the $e^+e^- \rightarrow h Z$ and $e^+e^- \rightarrow h A$
reactions.  We implement these constraints as upper limits on the
number of events seen in the four LEP experiments, as reported
in~\cite{LEP2}.  To do this, we first calculate the $h Z$ and $h A$ cross
sections
including initial-state radiation effects, then multiply by the
luminosities and divide by the detection efficiencies quoted by the
four collaborations in the different search modes, to obtain the total
number of events expected in all the LEP experiments. We then compare
with the reported results of the four collaborations~\cite{LEP2},
including
the announced candidates in the different channels and taking account of
their reported masses and resolution errors \cite{gl}. 
Since we include the full
renormalization-group-improved mass formulae \cite{hhh} for the Higgs
boson masses, which are sensitive to the stop mass spectrum, we also
implement the latest available constraints on stop production at
LEP~2~\cite{stopLEP}.

\section{Parameter Constraints}

We start by using the upper limits on $\sigma_{+-}$ and
selectron production to derive a joint constraint in the $(\m12,m_0)$
plane, assuming a universal input soft supersymmetry-breaking mass
$m_0$ for the the left and right sleptons. We do this by first fixing the
value of
$\tan\beta$, then, for each value of $m_0$, varying $\mu$ and plotting
the minimum value of $m_{1/2}$ for which both the $\sigma_{+-}$ and
selectron constraints are respected. This provides the 
boundary of the hatched excluded domain, labelled ``LEP", shown in
Fig.~2 for negative $\mu$ and in Fig.~3 for positive $\mu$, for
representative choices of tan$\beta$. We focus attention on the central
tan$\beta =2$ cases shown in panels (a,b) of these figures, with
outlying cases tan$\beta = \sqrt{2}, 35$ shown in panels (c,d).
We also
indicate by diagonal hatching in Figs.~2 and 3 the domains of the
$(\m12,m_0)$
plane which are excluded theoretically in this framework because the
lightest neutralino $\chi$ is heavier than the lighter stau
${\tilde \tau}_R$.  It is apparent that the hatched LEP
exclusion domain provides a
non-trivial lower bound on $m_{\chi}$, even in the absence of any
further theoretical input.  Although the LEP curve is somewhat
re-entrant when $m_0 \sim 70$ GeV, it is bounded well away from the
$m_{1/2} = 0$ axis. Thus, the previous loophole in the ALEPH
analysis~\cite{achi}
which allowed $m_{\chi} \sim 0$ in the neighbourhood of tan$\beta =
\sqrt{2}$ is now excluded~\footnote{This possibility may also be
constrained by searches for $W^{\pm}$ decays into charginos and
neutralinos: see~\cite{KZ}.}, as is the other previous loophole at large
$m_0$ for tan$\beta \sim 1.01$. The resulting experimental lower limit
on $m_{\chi}$ as a function of tan$\beta$ is shown by the curve labelled ``LEP'' in
Fig.~1.

The constraints in the
$\m12, m_0$ plane obtainable from squark and gluino searches at
Fermilab~\cite{gluino},
assuming universality for the gaugino masses, are
essentially the same as recorded in Figs.~1 and 3
of our LEP~1.5 analysis~\cite{efos}. Therefore, for reasons of simplicity,
we have not noted them explicitly in Figs.~1 and 3 of this paper.
However, they are shown for reference
in Fig.~2b, where it is seen that
they again play the valuable r\^ole
of excluding parts
of the re-entrant regions in Figs.~2 and 3, though this r\^ole
is less important here than in the LEP~1.5 analysis.

Further interesting constraints on the supersymmetric parameter
space may be obtained by taking account of
the LEP constraints on supersymmetric Higgs bosons~\cite{LEP2}. In the
absence of
further theoretical input, one must allow arbitrarily large values of
$m_A$, rendering the $hA$ search irrelevant and retaining just the
$hZ$ search.  The tree-level Higgs mass asymptotes to $m_Z
|\cos{2\beta}|$ for $m_A\gg m_Z$, and for small $\tan\beta$ this lies
well below the experimental lower bound.  However, the
renormalization-group-improved formula for $m_h$~\cite{hhh} depends on the
sfermion masses, in particular the stop and (for very large
$\tan{\beta}$) the sbottom masses, and the constraints on the Higgs
mass coming from the $e^+e^- \rightarrow h Z$ searches can be
satisfied even for low $\tan{\beta}$ if the sfermion masses are
sufficiently large. We henceforth assume that the
input values of the soft supersymmetry-breaking 
squark masses are also equal to $m_0$. 
The low-scale renormalized sfermion masses are given by $m_{\tilde
f}^2=m_0^2+C_f\m12^2+
m_f^2+O(M_Z^2)$, where the contributions $\propto\m12^2$ are due to
the renormalization group evolution of the soft masses from $M_X$
to $M_Z$~\cite{mssm}. 
The Higgs search bound can then be translated into a
contour in the $(\m12,m_0)$ plane, restricting one to large $\m12$
and/or $m_0$.  Since the radiative corrections to $m_h$ are only
logarithmically sensitive to the sfermion masses, the bounds on
$\m12$ and $m_0$ increase rapidly as $\tan\beta$ becomes small.
The Higgs mass can be increased by introducing mixing between stop
eigenstates, although this is constrained by current lower limits on
the mass of the lightest stop \cite{stopLEP}, whilst Higgsino loops can
provide a small negative contribution to $m_h$ \cite{hhh}~\footnote{The
corrections also depend upon the top mass, which we take here to have
the central value of $171\gev$.}.

It is important to note that the extension of the universality
assumption to the input stop and sbottom masses constrains the
allowed values of other supersymmetric model parameters. Consider first
the renormalization group evolution of $A_t$ down from the unification
scale. For $A_t$
much larger than $\m12$, the leading-order running of $A_t$ is given by
$dA_t/dt\approx 6 \lambda_t^2 A_t/8\pi^2$~\cite{mssm}, 
so that (for constant $\lambda_t$) $A_t$ decreases as a
power law with scale, with an exponent that is roughly 1/12 for
$\lambda_t \sim 1$. Thus, $A_t$ is reduced by an order of magnitude in
its evolution from $M_X$ to $M_Z$, and large
$A_t(M_Z)$ requires extremely large $A_t(M_X)$ if $\m12$ is small.  On the
other hand,
$A_t$ itself enters into the running of other soft masses, in
particular the stop mass-squared parameters, and extreme values for $A_t$
drive the stop masses negative at $M_Z$.  In practice, we find that
$0\le A_t \le 500 \gev$ covers the allowed range of $A_t$ for the
relevant values of $\m12$, with the positive gaugino mass contribution to
the running of $A_t$ driving $A_t>0$ even for $A_t(M_X)<0$.  Since the
bottom Yukawa is $\ll 1$ for moderate $\tan\beta$, $A_b$ is not
so tightly constrained, and we allow $-2\tev\le A_b \le2\tev$, to
include all values of $A_b$ which do not lead to charge and/or color
breaking in the scalar sector~\cite{ccb}.  With the parameters $A_t$
and $A_b$ bounded as above, stop and (for large $\tan\beta$) sbottom
mixing now limit the allowed range for $\mu$, since a large 
value of $|\mu|$ may push
the physical mass of the lightest stop (sbottom) below its current
experimental limit~\cite{stopLEP}.  At very large $\tan\beta$, where
$\lambda_b$ is not small, $A_b$ is further restricted, but the range
of $\mu$ allowed by sbottom mixing is quite insensitive to the limits
on $A_b$ in this case.

To establish the bound in the $(\m12,m_0)$ plane coming from Higgs
searches, we vary $\mu, A_t$ and $A_b$ over the range allowed by the
stop and sbottom mass limits, the absence of charge and
colour-breaking minima in the scalar potential~\cite{ccb}, and the
renormalization group evolution of the trilinear couplings. 
We find that only the regions
bounded by the curves labelled ``Higgs'' in Figs.~2(a,c) and 3(a,c) 
permit solutions
with sufficiently small Higgs production rates.  The curves bend to
the left at large $m_0$, where large sfermion masses lead to greater
positive radiative corrections to the Higgs mass, and the 
Higgs curve strikes the chargino bound
at sufficiently large $m_0$.
For small tan$\beta$, where the tree-level Higgs masses are small, the
Higgs search constraints provide stronger limits on $m_{\chi}$ than
those obtained from the chargino and slepton searches alone, as seen
in Figs.~2 and 3.  The
narrow dips at $m_0\sim 70\gev$ are excluded~\footnote{As already
mentioned, this r\^ole is also played by the 
Fermilab gluino searches~\cite{gluino}.}, and the
smallest neutralino masses come from points along the chargino bound
at large $m_0$. 
This effect is carried over to Fig.~1, where the
branch labelled ``H''
is the lower bound on $m_{\chi}$ due to combining the
LEP~2 searches for sparticles and Higgs bosons.  For large
$\tan\beta$, the tree-level Higgs mass is already large enough to
yield a sufficiently small Higgs production rate (recall that we are free
to choose a large value for $m_A$), and the Higgs
searches provide no additional bound on the $(\m12,m_0)$ plane, hence
the absence of a ``Higgs'' curve in Fig.  2d and 3d.  
The Higgs searches
fail to improve on the constraints coming from chargino, neutralino
and slepton searches alone for all $\tan\beta\ga 2.8$ for $\mu<0$ and
$\tan\beta\ga 1.7$ for $\mu>0$.

\section{Incorporation of Cosmological Constraints on Neutralino Dark
Matter}

We now combine these accelerator bounds with
cosmological bounds, assuming that the lightest neutralino is
the lightest supersymmetric particle, and is absolutely
stable, as in models with a conserved $R$ parity and
a relatively heavy gravitino~\cite{ehnos}.

It is well known that  in a considerable domain of the supersymmetric
parameter space the neutralino is an interesting dark matter candidate.
In what follows, we focus on region of the parameter space in which the
relic abundance of neutralinos left over from annihilations in the early 
Universe contributes a significant though not excessive amount to the
overall energy density. Denoting by $\Omega_\chi$ the fraction of the
critical energy density provided by
neutralinos, we focus on the region of parameter
space in which
\beq 
0.1 \le \Omega_\chi h^2 \le 0.3
\label{orange}
\eeq
The lower limit in eq.(\ref{orange}) is motivated by astrophysical
relevance. For lower values of $\Omega_\chi h^2$, there is not enough
neutralino dark matter to play a significant r\^ole in structure
formation, or constitute a large fraction of the critical density.
Regions of parameter space in which the neutralino density fall short
of this bound are not excluded, they are simply not of cosmological
interest.  In Figs.~2 and 3, only the light-shaded regions admit a neutralino
with a relic density $\ohsq>0.1$.

The upper bound in (\ref{orange}), on the other hand, is an absolute
constraint,
derivable from the age of the Universe, which can be expressed as
\beq
H_0 t_0 = \int_0^1 dx \left(1 - \Omega - \Omega_\Lambda + \Omega_\Lambda
x^2
 + \Omega /x\right)^{-1/2}
\label{ht}
\eeq
In (\ref{ht}), $\Omega$ is the density of matter relative to 
critical density, while $\Omega_\Lambda$ is the equivalent contribution
due a cosmological constant.
Given a lower bound on the age of the Universe, one can establish an
upper bound on $\Omega h^2$ from eq.(\ref{ht}).
In light of the new Hipparcos data~\cite{Hipparcos}, a safe lower bound to
the age of the Universe
is $t_0 \ga 12$ Gyr, which translates into the upper bound given in
(\ref{orange}).
This bound is independent of the value of $\Omega$ or $\Omega_\Lambda$, so
long as $\Omega + \Omega_\Lambda \le 1$.

Two generic possibilities for the composition of a possible
neutralino dark matter candidate should be distinguished~\cite{ehnos}: it
may
have mainly a gaugino composition, in which case its mass is more 
sensitive to $\m12$ than to $\mu$, or it may be mainly a
higgsino~\cite{higgsino},
in which case its mass is more sensitive to $\mu$. 
Much of the higgsino region has now been excluded by 
LEP~2~\footnote{Moreover, this higgsino
region is not accessible if dynamical EWSB is implemented with
universal input parameters, as discussed below.}. This is
because neutral higgsino dark matter
particles should weigh less than 80 GeV, since heavier higgsinos
annihilate
rapidly into $W^+W^-$, suppressing the relic density below the
relevant range (\ref{orange}) \cite{osi3}. 
On the other hand, since $m_{\chi^{\pm}} - m_{\chi}$
is small in the higgsino region, the LEP 2 chargino searches now
effectively exclude $m_{\chi} \la 75$ GeV, leaving a narrow allowed range
75 GeV $\la m_{\chi} \la$ 80 GeV.
The fate of this remaining region will soon be
decided by higher-energy runs of
LEP~2: among other searches, these should be able to probe $m_{\chi^{\pm}}
\la 95$ GeV for $m_{\chi^{\pm}} - m_{\chi} \ga 5$ GeV, sufficient to
discover or exclude a higgsino dark matter candidate.

An important consequence of the upper limit on $\ohsq$ is the
exclusion of a large region in $m_0$ for at least some range of values
of $m_{1/2}$, which results from combining cosmology with the LEP
supersymmetric Higgs constraint.  Gaugino-type neutralinos annihilate
in the early universe predominantly through sfermion exchange into
fermion pairs.  Large sfermion masses shut off this annihilation
channel and lead to large neutralino relic densities, in violation of
the upper limit in (\ref{orange}).  Since the sfermion masses depend
on both $m_0$ and $\m12$ 
via the renormalization group equations, (\ref{orange}) therefore
places an upper bound on $m_0$ and $\m12$ for gaugino-type
neutralinos. In our previous analysis~\cite{efos},
the relic density
could have been reduced to an acceptably low value, even for
arbitrarily large values of $m_0$, by choosing a small value of
$|\mu|$, which causes the lightest neutralino to become a
gaugino/higgsino mixture.  Including Higgs production constraints
removes this freedom, as regions of low $\m12$ yield too low a
Higgs mass unless $\mu$ is taken to be very large.  As described
above, this is particularly important at low $\tan\beta$, where the
tree-level Higgs masses are small. The result of combining the LEP
Higgs constraints with $\ohsq<0.3$ is shown as the dashed line in
Figs.~2(a,c) and 3(a,c).  In Fig.~2a, the dashed line turns up at $\m12\ga
155\gev$, and
at $\m12\ga 164\gev$ for $\mu > 0$ in Fig.~3a,
where the Higgs is sufficiently heavy for low $|\mu|$.  
In Fig~2a, the gap at $\m12\sim 85\gev$ is due to the 
presence of a Higgs pole in the neutralino annihation cross-section.
Similarly the gaps at
$\m12\sim 95\gev$ and $\m12\sim 115\gev$ are  
due to the presence of a Higgs  and $Z^0$ pole respectively in Fig.~3a. 
 For $\tan\beta=35$, no additional
limit is obtained by combining the Higgs and cosmological
constraints.  

Of course an alternate way of satisfying the Higgs bound is to take
$m_0$ very large, rather than $\mu$:  taking $\mu$ sufficiently small
then yields again a mixed neutralino for which the upper limit of
(\ref{orange}) is easily satisfied.  The dashed curves thus bend back
to the left at very large $m_0$ and strike the chargino bound.  This
intersection occurs at values of $m_0$ well above the range plotted:
for $\mu<0$ this occurs at $m_0\sim 700\gev$ for
$\tan\beta=2$, and at $m_0\sim 2\tev$ for $\tan\beta=\sqrt{2}$.
For $\mu<0$, and for low $\tan\beta$, these large values of $m_0$
permit the lowest values of $\m12$ and the lightest neutralinos.  In
Fig.~1, the branch labelled ``C'' is the lower bound on $\mchi$ coming
from combining the LEP experimental limits with the cosmological
constraint $\ohsq<0.3$.  Comparing curves H and C in
Fig.~1, we see that the additional cosmological constraint is in this
case only relevant for low $\tan\beta$.  The difference in bounds is
due to the requirement that $\mu$ be sufficiently small to yield a
mixed neutralino.  If one were to require that $m_0<500\gev$, the
cosmological constraint would yield much tighter bounds on
$\mchi$ at low $\tan\beta$. For $\mu>0$, a
$|\mu|$ sufficiently small to yield a mixed neutralino also gives a
small chargino mass, and in contrast to the $\mu<0$ case, the lowest
neutralino masses at low $\tan\beta$ come from the corner between the
Higgs bound and the $\ohsq=0.3$ contour visible on Fig. 3a.  This
explains why branch C is significantly higher in Fig.~1b than in
Fig.~1a at low $\tan\beta$.

\section{Implications of Universal Masses for Higgs Scalars}

We now supplement the above
experimental and cosmological considerations by
extending the theoretical assumption of GUT-scale universality for
the input scalar soft supersymmetry-breaking
masses to those in the Higgs sector. In this case, renormalization-group
calculations leading to dynamical EWSB with the correct Higgs
vacuum expectation values fix the previously undetermined
parameters $|\mu|$ and $m_A$, for any given choice
of values of $m_0, m_{1/2}$~\footnote{The assumed values of other
MSSM model parameters such as the trilinear soft supersymmetry-breaking
parameters $A$ are not essential for this argument. In order to
implement dynamical EWSB down to the smallest possible value of 
tan$\beta \sim 1.2$ as seen in Fig.~1, for this analysis we allow the top
mass to be as low as
161~GeV.}.
These restrictions have the effect of further
strengthening the above lower limits on $m_{\chi}$. First, one is no
longer permitted to vary $\mu$ to find the lowest possible rates of
chargino and selectron production, so that the boundary of the previous
hatched LEP exclusion domain in Figs.~2 and 3 is moved to the right,
as shown.
This change is least important for intermediate values of
tan$\beta \sim 3$, but always tends to fill in the
previously re-entrant portion of the LEP region.
 
The most dramatic effect of the scalar-mass universality assumption
appears when it is implemented for the limits coming from the searches
for Higgs bosons. It both
constrains the renormalization-group-improved calculation of $m_h$ and
enables the $hA$ search to come into play.  The former effect causes
the $hZ$ search to provide a very important lower limit on $m_{1/2}$,
particularly for small tan$\beta$. This effect is so significant for
tan$\beta = \sqrt{2}$ that for $\mu < 0$ it 
requires $m_{1/2} \ga 1$ TeV, far to the
right of the corresponding panel in Fig.~2c. The Higgs lower bound with
universal input scalar masses is visible  in the tan$\beta = 2$
panel of Figs.~2 and 3 labelled ``UHM'' (for universal Higgs mass).
For $\mu<0$ it becomes of comparable significance to the other
bounds when $\tan\beta \sim 3$, and for slightly smaller $\tan\beta$
when $\mu>0$.  The second constraint due to the
fixing of $m_A$ becomes significant for large values of tan$\beta \ga
35$, which is of relevance to models in which Yukawa couplings are
also assumed to be universal. All of the Higgs bounds bend to the
left at sufficiently large $m_0$. For example, the Higgs curve in
Fig.~2b strikes the chargino bound at $m_0\ga 800\gev$.  Thus in the
absence of an independent constraint on $m_0$ (see below), the Higgs
bound at low $\tan\beta$ is effectively the chargino bound at large
$m_0$.  The resulting strengthened lower bound on $m_{\chi}$ is shown
as the thin solid line labelled ``UHM'' in Fig.~1.

\section{Combining Cosmology and Universality}

After applying separately the cosmological and universality
constraints in the two previous sections, we now apply them both
simultaneously.  A first comment is that higgsino dark matter is not
compatible with scalar-mass universality, since dynamical EWSB then
fixes $|\mu|$ so that the lightest neutralino is in the gaugino
region.  The dark shading in Figs.~2 and 3 delimits the region in which the
neutralino relic density satisfies (\ref{orange}), and we see that the
universality assumption restricts significantly the region of the
$m_{1/2}, m_0$ plane in which the relic density lies within the
favoured range. Indeed, for tan$\beta = 35$, the favoured region lies
at larger values of $\m12$ not shown.
One sees the effects of enhanced annihilation
rates through adjacent Higgs and $Z$ poles, which create low-density
channels through, and distortions of, the favoured dark-shaded region.
In Fig.~2 these are shown in their entirety, but for reasons of
clarity in Fig.~3 they are shown only in the
regions not already excluded by LEP.

In Fig.~1, the thick branch of the solid UHM line
labelled ``cosmo + UHM'' describes the improvement to the
universal scalar mass $\mchi$ bound at low $\tan\beta$ due to the true
cosmological constraint $\ohsq<0.3$.  The thick branch of the solid line
labelled ``DM + UHM'' shows
the improvement at high $\tan\beta$ from including the preference that
$\ohsq>0.1$.  The latter provides a significant improvement in
the bound, amounting to almost a factor of two at high $\tan\beta$
over the solid grey $\mchi$ bound from the previous section.  As in
\cite{efos,efosp}, the kink around tan$\beta \ga 3$ and the
tight constraint on $\mchi$ at large $\tan\beta$ for $\mu > 0$ arise
from the necessity of being to the right side of the $Z$ pole in
order to have a sufficiently high relic density compatible with
(\ref{orange}). 

Perhaps the most dramatic impact, however, is seen at low tan$\beta$,
where we have already noted that the LEP Higgs constraint imposes a
very strong constraint on the $(m_{1/2},m_0)$ plane if universal
scalar masses are assumed.  The cosmological bound on the sfermion
masses now forbids the large values of $m_0$ which 
previously permitted low
$\m12$, and consequently relatively low $\mchi$, at low $\tan\beta$.
As noted above, the Higgs exclusion curve moves rapidly to larger
$\m12$ as $\tan\beta$ is decreased, and the improvement in the $\mchi$
bound at low $\tan\beta$ is commensurately rapid. This explains
the different analytic forms of the constraints on $\mchi$ for tan$\beta
\la 2.6$ for $\mu < 0$ and tan$\beta \la 1.8$ for $\mu > 0$.

This is particularly important because there is an
{\em upper} bound on the value of $m_{1/2}$ for which the cosmological
relic density of neutralinos can be kept within the cosmologically
interesting range (\ref{orange})~\cite{upper}, if the universality
assumption is
extended to Higgs mass parameters, which implies that the lightest
neutralino is gaugino-like~\footnote{We emphasize that
  this bound does not apply to the higgsino-like case, which is
  possible if Higgs universality is relaxed: see for
  example~\cite{EG}.}. As shown in Fig~3(c), for low enough $\tan\beta$,
the Higgs
bound moves entirely to the right of the dark-shaded region, and for
$\mu<0$, the cosmologically allowed range with $\ohsq<0.3$ is actually
{\em incompatible} with the Higgs lower limit on $m_{1/2}$ for
tan$\beta \la 1.7$. This cosmological upper bound on $\m12$ varies
only weakly for tan$\beta \la 2$. We conclude
that there is no range of $m_{1/2}$
compatible with all the constraints provided by the LEP experiments,
the upper bound on the cosmological relic density, and the theoretical
assumption of scalar-mass universality, for sufficiently small
tan$\beta \la 1.7$. Hence, {\em there is a lower bound}
\begin{equation}
\hbox{tan}\beta \ga 1.7
\label{nosolution}
\end{equation}
if all these constraints are applied simultaneously.  
Similarly, for $\mu > 0$, the bound is $\tan\beta~\ga~1.4$.

We emphasize that this bound comes from merely imposing an upper bound
on the relic density, which is simply due to the lower limit on the
age of the universe of 12 Gyr: the constraint (\ref{nosolution}) does
{\it not} require that $\ohsq>0.1$.  We also note that, due to the
sensitivity to $\tan\beta$ of the Higgs bound on $\m12$, this bound is
quite robust.  The dependence of the bound (\ref{nosolution}) on such
input parameters as $A_t(M_X)$ and $m_t$, as well as any
residual uncertainty in the extraction of the Higgs mass, can be
parameterized in terms of their effect on $m_h$: any change which
produces a $1\gev$ increase in the Higgs mass will decrease the bound
(\ref{nosolution}) on $\tan\beta$ only by roughly $0.01$.  As dicussed
in Section 3, the value of $A_t(M_Z)$ is relatively insensitive to
$A_t(M_X)$, particularly at low $\tan\beta$, where the top Yukawa
becomes quite large as it is run to $M_X$.  Therefore the uncertainty
in $m_h$ near the limit (\ref{nosolution}) due to changes in
$A_t(M_X)$ is negligible, though the radiative corrections to $m_h$ do
increase with $m_t$.  However, for $m_t$ too large,
the running of the top Yukawa becomes non-perturbative below $M_X$.  
The upper limit this imposes on $m_t$ decreases as $\tan\beta$ becomes
small, and at low $\tan\beta$ (i.e., near the bound
(\ref{nosolution})) we use the largest $m_t \sim 161$~GeV for which the
running of the top Yukawa remains perturbative up to $M_X$.  Therefore
variations in $m_t$ will not decrease the bound
(\ref{nosolution}), although they can increase it for smaller $m_{t}$.

\section{Conclusions and Prospects}

We have seen in this paper how the recent higher-energy LEP~2
data~\cite{LEP2W,LEP2} impose interesting new constraints on the
parameter space of the MSSM, in particular on the mass of the lightest
neutralino $\chi$, assuming that it is stable. Direct searches indicate
that $m_{\chi} \ga 17$ GeV, and exclude a large fraction of the domain of
MSSM parameter space where the lightest neutralino is
Higgsino-like~\footnote{As already noted, future LEP~2 runs should
determine the fate of this option.}.
The absolute lower bound on $m_{\chi}$
is increased to 40 GeV if
it is assumed that all the soft supersymmetry-breaking scalar (slepton,
squark and Higgs) masses are universal at some GUT input scale, and that
the relic neutralino density fall within the range (\ref{orange}).
Moreover, these assumptions are incompatible with the LEP~2 limits 
unless tan$\beta \ga 1.7$ for $\mu<0$.

In addition to their implications for dark matter detection strategies,
the LEP~2 limits are beginning to raise questions for supersymmetric
model builders. Models which incorporate Yukawa unification as well as 
the the universal scalar masses invoked here tend to favour values of
tan$\beta \sim 1.8$ or $56$~[see~\cite{CK} and references therein]. The
former option is already strongly
constrained by LEP~2, and would become untenable if the further upgrades
of the LEP energy that are foreseen fail to reveal a supersymmetric
Higgs boson. The combination of this with other LEP~2 searches would
have sensitivity to tan$\beta \la 3$ for $\mu<0$. Thus model-builders
may soon have to envisage the relaxation of at least one of the GUT
universality and unification assumptions that are conventionally made in
constraining the parameters of the MSSM~\cite{cmssm}.

\vskip 0.5in
\vbox{
\noindent{ {\bf Acknowledgments} } \\
\noindent  J.E. would like to thank the University of Minnesota
for kind hospitality while parts of this work were being done.
M.S. is grateful for the use of codes written by Gerardo Ganis.
This work was supported in part by DOE grant DE--FG02--94ER--40823.}

%
%--------------------------------------------------
\begin{figure}
\begin{center}
\mbox{\epsfig{file=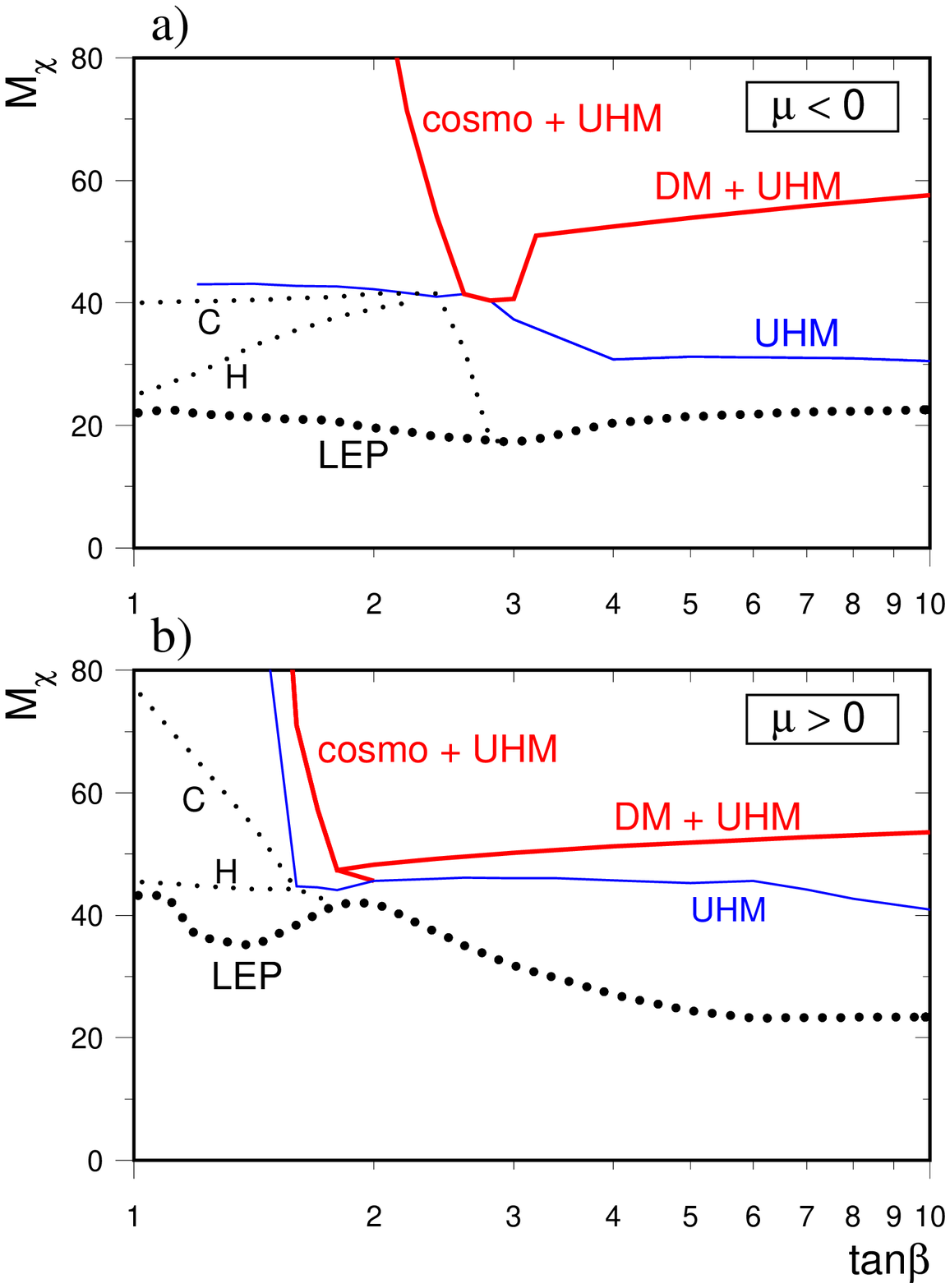,height=16.5cm%
,bbllx=10mm,bblly=10mm,bburx=170mm,bbury=220mm}}
\end{center}
\caption[.]{ 
  Various lower limits on $m_{\chi}$ (in GeV) 
  obtained using different
  experimental and theoretical inputs are compared, as functions of
  $\tan\beta$, for both (a)~$\mu < 0$ and (b)~$\mu > 0$.  The
  dotted lines labelled ``LEP'' are obtained by combining the
  unsuccessful LEP~2 searches for charginos and selectrons, allowing
  $\mu$ to vary over the range allowed by the bounds on $A_t$
  discussed in the text.
  The dotted branches labelled ``H'' and ``C'' additionally
  incorporate the requirements that
  lightest neutral supersymmetric Higgs boson not be seen at LEP, and
  also the relic cosmological density $\ohsq<0.3$, respectively.
  The solid lines are bounds incorporating the theoretical
  assumption of universal scalar masses as GUT inputs into dynamical
  calculations of the electroweak symmetry-breaking scale. The solid
  lines include the LEP experimental searches for charginos,
  selectrons and Higgs bosons, with the branches ``cosmo + UHM'' and
  ``DM + UHM''
  incorporating the constraints $\ohsq<0.3$ and  $0.1<\ohsq<0.3$,
  respectively. 
\label{mchi_tb} }
\end{figure}
%--------------------------------------------------
%
%--------------------------------------------------
\begin{figure}
\begin{center}
\mbox{\epsfig{file=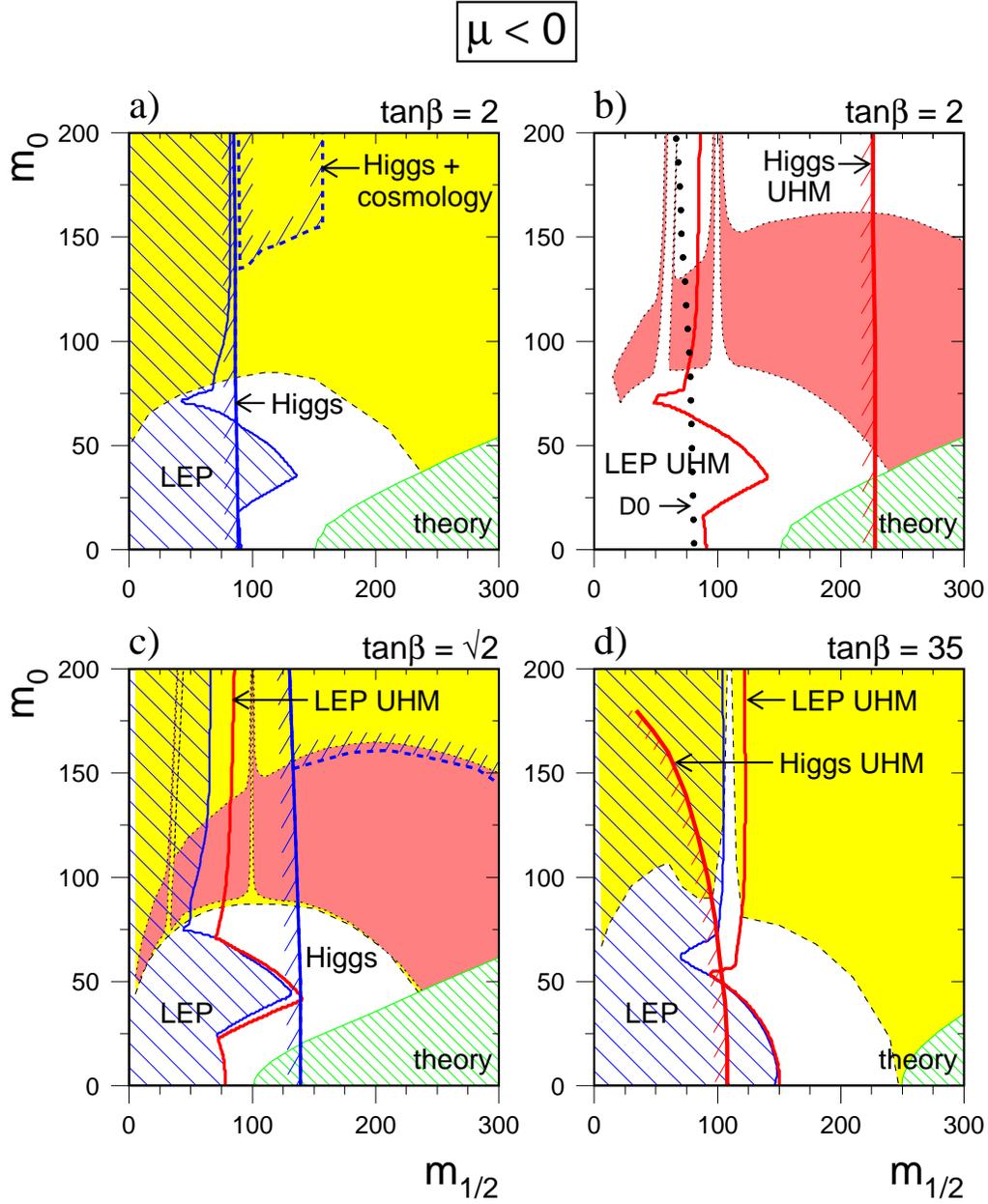,height=16cm%
,bbllx=10mm,bblly=40mm,bburx=170mm,bbury=235mm}}
\end{center}
\caption[.]{ 
  We display for $\mu < 0$ and (a,b)~$\tan\beta = 2$,
  (c)~$\sqrt{2}$, (d)~$35$, the domains of the ($m_{1/2}$,~$m_0$)
  plane (in~GeV)
  that are excluded by the LEP~2 chargino and selectron searches, 
  both without (hatched) and with the assumption of Higgs
  scalar-mass universality. We also display 
  the domains that are excluded by Higgs
  searches (solid lines) without (a,c) and with (b,d)
  the assumption of universal scalar masses for Higgs bosons (UHM).
  Also shown are the regions that are
  excluded cosmologically because $m_{{\tilde \tau}_R} < m_{\chi}$,
  and the domains that have relic neutralino densities in the favoured
  range (\ref{orange}) with (dark) and without (light shaded) the
  scalar-mass universality assumption.  For clarity, for the case
  $\tan\beta=2$ we display separately the bounds without and with the
  assumption of Higgs scalar-mass universality in Figs.~2b and~2c
  respectively. 
\label{muneg} }
\end{figure}
%--------------------------------------------------
%
%--------------------------------------------------
\begin{figure}
\begin{center}
\mbox{\epsfig{file=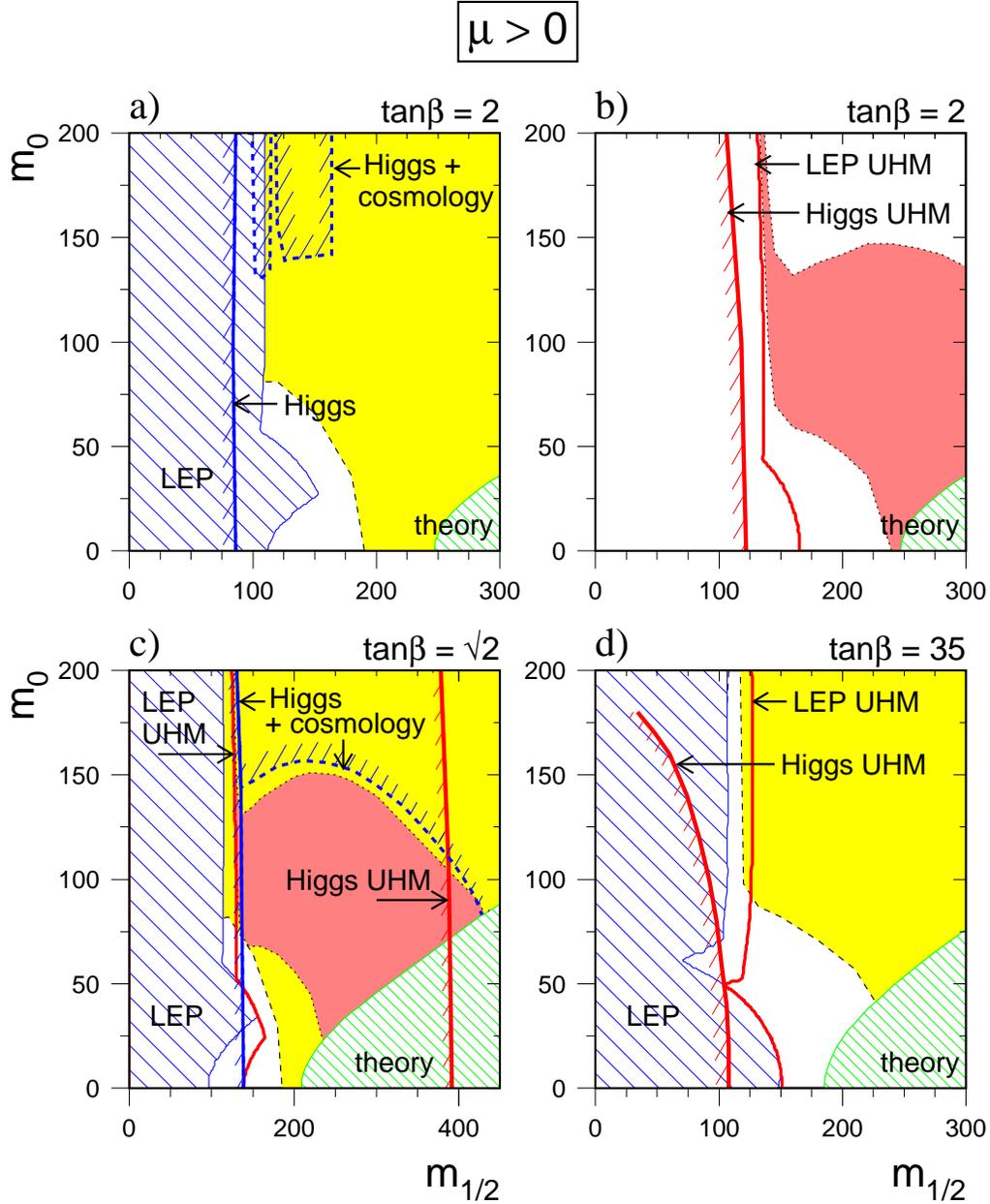,height=16cm%
,bbllx=10mm,bblly=40mm,bburx=170mm,bbury=235mm}}
\end{center}
\caption[.]{ 
  As Fig.~2, but for $\mu>0$, and with the cosmological
  density contours suppressed in the domains excluded by LEP~2 searches.
  Note the different horizontal scale in panel~(c), chosen to exhibit
  the cosmological upper limit on $\m12$ and its interplay with the
  assumption of universal scalar masses for Higgs bosons (UHM).
\label{mupos} }
\end{figure}
%--------------------------------------------------
 
\end{document}